\documentclass[fleqn,11pt,twoside]{article}

\usepackage{amsthm,amsthm,amssymb, color, xcolor,epsfig, graphics, subfigure}

\usepackage{amsmath, graphicx, latexsym, lscape}

\makeatletter
\newcommand{\copyrightnote}[2]{{\renewcommand{\thefootnote}{}
 \footnotetext{\small\it
\begin{flushleft}
 \copyright \ #1   #2  
\end{flushleft}}}}

\newcommand{\Name}[1]{\begin{flushleft}
                       \LARGE \bf #1
                       \end{flushleft}\vspace{-3mm}}

\newcommand{\Author}[1]{\begin{flushleft}
                       \it #1 \end{flushleft}}

\newcommand{\Address}[1]{\begin{flushleft}
                       \it #1 \end{flushleft}}

\newcommand{\Date}[1]{\begin{flushleft}
                      \small  \it #1 \end{flushleft}}

\newcommand{\evenhead}{Author \ name}
\newcommand{\oddhead}{Article \ name}

\renewcommand{\@evenhead}{
\hspace*{-3pt}\raisebox{-15pt}[\headheight][0pt]{\vbox{\hbox to \textwidth
{\thepage \hfil \evenhead}\vskip4pt \hrule}}}
\renewcommand{\@oddhead}{
\hspace*{-3pt}\raisebox{-15pt}[\headheight][0pt]{\vbox{\hbox to \textwidth
{\oddhead \hfil \thepage}\vskip4pt\hrule}}}
\renewcommand{\@evenfoot}{}
\renewcommand{\@oddfoot}{}

\long\def\@makecaption#1#2{%
  \vskip\abovecaptionskip
  \sbox\@tempboxa{\small \textbf{#1.}\ \ #2}%
  \ifdim \wd\@tempboxa >\hsize
    {\small \textbf{#1.}\ \ #2}\par
  \else
    \global \@minipagefalse
    \hb@xt@\hsize{\hfil\box\@tempboxa\hfil}%
  \fi
  \vskip\belowcaptionskip}

\newcommand{\JNMPnumberwithin}[3][\arabic]{%
  \@ifundefined{c@#2}{\@nocounterr{#2}}{%
    \@ifundefined{c@#3}{\@nocnterr{#3}}{%
      \@addtoreset{#2}{#3}%
      \@xp\xdef\csname the#2\endcsname{%
        \@xp\@nx\csname the#3\endcsname .\@nx#1{#2}}}}%
}

\newcommand{\resetfootnoterule} {
  \renewcommand\footnoterule{%
  \kern-3\p@
  \hrule\@width.4\columnwidth
  \kern2.6\p@}
}

\renewcommand{\footnoterule}{}

\makeatother

\theoremstyle{definition}

\begin{document}

\renewcommand{\evenhead}{ {\LARGE\textcolor{blue!10!black!40!green}{{\sf \ \ \ ]ocnmp[}}}\strut\hfill 
Laurent Delisle and Amine Jaouadi
}
\renewcommand{\oddhead}{ {\LARGE\textcolor{blue!10!black!40!green}{{\sf ]ocnmp[}}}\ \ \ \ \  
A vector bilinear framework for soliton dynamics
}

\thispagestyle{empty}
\newcommand{\FistPageHead}[3]{
\begin{flushleft}
\raisebox{8mm}[0pt][0pt]
{\footnotesize \sf
\parbox{150mm}{{\textcolor{blue!10!black!40!green}{{\bf Open Communications in Nonlinear Mathematical Physics}}}
\ \ {Special Issue: Hietarinta}, 2026\\[0.1cm]
\strut\hfill 
ocnmp:18025
pp #2\hfill {\sc #3}}}\vspace{-13mm}
\end{flushleft}}

\FistPageHead{1}{\pageref{firstpage}--\pageref{lastpage}}{ \ \ }

\strut\hfill

\strut\hfill

\copyrightnote{The author. Distributed under a Creative Commons Attribution 4.0 International License}

\begin{center}

{\bf {\large A Special OCNMP Issue in Honour of Jarmo Hietarinta}}\\[0.2cm]
{\bf {\large on the Occasion of his 80th Birthday}}
\end{center}

\smallskip

\Name{A Vector Bilinear Framework for Soliton Dynamics in Coupled Modified KdV Systems}

\Author{Laurent Delisle and Amine Jaouadi}

\Address{LyRIDS, ECE Engineering School, OMNES Education, 10 rue Sextius Michel, 75015 Paris - France }

\Date{Received April 15, 2026; Accepted May 6, 2026}

\setcounter{equation}{0}

\smallskip

\noindent
{\bf Citation format for this Article:}\newline
Laurent Delisle and Amine Jaouadi,
A vector bilinear framework for soliton dynamics in coupled modified KdV systems,
{\it Open Commun. Nonlinear Math. Phys.}, Special Issue:\,Hietarinta, ocnmp:18025, \pageref{firstpage}--\pageref{lastpage}, 2026.

\strut\hfill

\noindent
{\bf The permanent Digital Object Identifier (DOI) for this Article:}\newline
{\it 10.46298/ocnmp.18025}
\strut\hfill

\begin{abstract}

\noindent 
We investigate the integrable structure and soliton dynamics of a coupled modified Korteweg--de Vries (cmKdV) system with a real symmetric coupling matrix. We introduce a vector reformulation of Hirota’s bilinear formalism in which both the bilinear equations and their solutions are expressed directly at the vector level, rather than through a component-wise construction. This formulation preserves the intrinsic structure of the coupled system and provides a compact framework for multi-component nonlinear wave dynamics. Within this approach, we construct explicit one-, two-, and three-soliton solutions in closed vector form and recover the three-soliton condition directly at the vector level, confirming consistency with integrability. The method enables a unified treatment of focusing, defocusing, and mixed-sign regimes. In particular, for indefinite coupling, it reveals the existence of nontrivial vector ground states, leading to soliton solutions on non-zero backgrounds. These results highlight the structural advantages of the vector bilinear approach and open perspectives for the study of more general nonlinear excitations in multi-component integrable systems.

\end{abstract}

\label{firstpage}


\section{Introduction}

Integrable nonlinear evolution equations have attracted sustainable interest in mathematical physics due to their rich analytical structure and the existence of exact solutions. Among these, soliton solutions constitute a remarkable class of nonlinear waves that preserve their shape, amplitude, and velocity after interactions. Traveling wave or one-soliton solutions arise in a wide range of integrable and non-integrable systems, while two-soliton solutions describe nonlinear interaction processes. The existence of exact three-soliton solutions is a much stronger property and is typically regarded as a hallmark of complete integrability. Over the years, several powerful analytical techniques have been developed to construct such solutions, including Bäcklund transformations \cite{Schiff}, Darboux transformations \cite{WXMa,Gu}, the inverse scattering transform (IST) \cite{Ablo,Tanaka,Wadati,Tsuchida}, and Hirota's bilinear formalism \cite{Hirota}. The latter provides a direct and constructive method based on rewriting nonlinear equations in bilinear form using Hirota derivatives, allowing systematic derivation of multi-soliton solutions through finite expansions. In particular, Hietarinta established that the existence of a three-soliton solution is equivalent to satisfying a strong integrability condition, and classified bilinear equations fulfilling this property for several important models \cite{Hietarinta1,Hietarinta2,Hietarinta3,Hietarinta4}.

Soliton solutions are not only of theoretical interest but also arise in a wide range of physical applications. Modified Korteweg--de Vries (mKdV)-type equations describe nonlinear wave propagation in nonlinear optics, including ultra-short pulse dynamics beyond the slowly varying envelope approximation \cite{Leblond,MaYL}. They also appear in fluid dynamics and rogue wave generation \cite{Slunyaev}, plasma and solid-state physics \cite{Khater,Singh,Verheest}, and traffic-flow models \cite{Ge,Song}. In multi-component media, vector solitons exhibit richer interaction dynamics due to the coupling between components, leading to complex behaviors such as energy exchange, symmetry-driven dynamics, and mixed bright-dark structures \cite{Ohta,Vijaya,Radha,Rad,Kanna1,Zhang}. Such effects are particularly relevant in Bose--Einstein condensates and reduced-dimensional quantum systems, where interactions and external potentials can be tuned experimentally. In this context, recent works have demonstrated analytical control of solitons' collisions \cite{DelisleDark}, symmetry-driven multi-soliton dynamics in reduced dimensions \cite{DelisleSym}, optically tuned soliton behavior in trapped condensates \cite{Celanie}, and applications in econophysics \cite{DelisleND}. These developments highlight the importance of analytical approaches capable of capturing the collective dynamics of coupled nonlinear systems.

The coupled modified Korteweg--de Vries (cmKdV) system provides a natural framework to investigate such multi-component nonlinear wave phenomena. Its complete integrability has been established using the inverse scattering transform \cite{Tsuchida}, and multi-soliton solutions have been constructed using Hirota-type methods \cite{Hirota1,Iwao}. However, in most existing approaches, the bilinearization procedure and the construction of solutions are carried out component-wise, treating each component of the vector solution separately. While this strategy yields explicit expressions, it tends to obscure the intrinsic vector structure of the system and the role of the coupling matrix in governing the collective nonlinear dynamics.



In this work, we introduce a vector reformulation of Hirota's bilinear formalism for the cmKdV system, in which both the bilinear equations and their soliton solutions are constructed directly in vector form. This approach can be viewed as a basis-independent extension of the standard Hirota method, where nonlinear interactions are encoded through quadratic forms involving the coupling matrix. As a result, the formulation preserves the native structure of the coupled system and provides a compact and structurally consistent framework for the analysis of multi-component soliton dynamics. In particular, it enables a unified treatment of focusing, defocusing, and mixed-sign regimes within a single formalism.

A key outcome of this vector framework is the natural emergence of nontrivial ground state solutions in the case of indefinite coupling matrices. This feature, which has no direct analogue in the scalar setting, leads to the construction of soliton solutions on non-zero backgrounds and reveals new classes of nonlinear excitations in multi-component systems. Within this formulation, we construct explicit one-, two-, and three-soliton solutions in closed vector form, and recover the three-soliton integrability condition directly at the vector level, thereby confirming that the proposed approach fully captures the integrable structure of the cmKdV system.

The paper is organized as follows. In Sec.~II, we introduce the coupled mKdV system and present its diagonalization. In Sec.~III, we develop the vector Hirota bilinear representation. In Sec.~IV, we construct vector multi-soliton solutions, including the explicit three-soliton case. In Sec.~V, we investigate one-soliton solutions on a nontrivial background. Finally, Sec.~VI contains concluding remarks and perspectives.

\section{The coupled modified Korteweg-de Vries system} To illustrate the vector generalization of the Hirota's bilinear formalism and its vector multi-soliton solution construction, we consider the cmKdV system.  These systems have been shown to be completely integrable using the IST \cite{Tanaka,Tsuchida}. Explicitly, the cmKdV systems reads as \cite{Tsuchida}:
\begin{equation}
    u_t+u_{xxx}+6(u^tAu)u_x=0,
    \label{cmKdV}
\end{equation}
where $A$ is a $N\times N$ real symmetric matrix and $u=u(x,t)$ is a real-valued $N-$component vector.  Subscripts means partial derivation with respect to the variable in question. For $N=1$, the cmKdV equations (\ref{cmKdV}) is a completely integrable generalization of the well known modified Korteweg-de Vries (mKdV) equation:
\begin{equation}
    u_t+u_{xxx}+6Au^2u_x=0,
\end{equation}
which is known to possess, in particular, multi-soliton solutions. Using the fact that the matrix $A$ is symmetric, we may write it as $A=P^tDP$ for $P\in O(N)$ an orthogonal matrix constructed from the eigenvectors of the matrix $A$ and $D$ is the diagonal matrix with diagonal entries corresponding to the eigenvalues of $A$. We assume, in this paper, that the eigenvalues are non zero which was assume to show the integrability of the cmKdV equations (\ref{cmKdV}) using the inverse scattering transform (IST). Setting $v=Pu$, the cmKdV system (\ref{cmKdV}) reduces to
\begin{equation}
    v_t+v_{xxx}+6(v^tDv)v_x=0.
    \label{V1cmKdV}
\end{equation}
The assumption $\mbox{det}(D)\neq0$ allows a rescaling of the system (\ref{V1cmKdV}). Indeed, if 
\begin{equation}
    D=\mbox{diag}(\lambda_1,\lambda_2,\cdots,\lambda_N)\quad \mbox{and}\quad v^t=(v_1\, \, v_2\,\,\,\cdots\,\, v_N),
\end{equation}
we may make the following substitutions:
\begin{equation}
    v_k=\frac{w_k}{\sqrt{\vert\lambda_k\vert}},
\end{equation}
for $k=1,2,\cdots, N$ in order to transform the coupled system of equations (\ref{V1cmKdV}) into :
\begin{equation}
    w_t+w_{xxx}+6(w^t\mathcal{E}w)w_x=0,
    \label{V2cmKdV}
\end{equation}
where $\mathcal{E}$ is a diagonal $N\times N$ matrix with diagonal entries equal $\pm 1$ depending on the sign of each eigenvalues. In particular, the focusing and defocusing cases correspond, respectively, to $\mathcal{E}=I_N$ and $\mathcal{E}=-I_N$ in equation (\ref{V2cmKdV}). The matrix $I_N$ representing the $N\times N$ identity matrix. The focusing case may occur when the matrix $A$ in equation (\ref{cmKdV}) is positive-definite. 

\section{Hirota's bilinear vector representation} We may obtain a Hirota bilinear representation of the cmKdV system (\ref{V2cmKdV}) using the transformation:
\begin{equation}
    w=\frac{F}{G},\label{subst}
\end{equation}
where $G=G(x,t)$ is a real-valued scalar function and $F=F(x,t)$ is a real-valued $N-$component vector. In order to do so, we need to define the Hirota derivative $\mathcal{D}$:
\begin{equation}
    \mathcal{D}_{x}^m\mathcal{D}_t^n(\mathcal{F\cdot\mathcal{G}})=(\partial_{x_1}-\partial_{x_2})^m(\partial_{t_1}-\partial_{t_2})^n\mathcal{F}(x_1,t_1)\mathcal{G}(x_2,t_2)\vert^{t_1=t_2=t}_{x_1=x_2=x,}\label{HirotaD}
\end{equation}
for $\mathcal{F}$ and $\mathcal{G}$ scalar functions. We need to generalize this Hirota derivative to vector functions. Let $e_j$ be the $j^{th}$ column of the unit $N\times N$ matrix and write
\begin{equation}
    F(x,t)=\sum_{k=1}^Nf_{k}(x,t)e_k,
\end{equation}
then we define
\begin{equation}
    \mathcal{P}(\mathcal{D}_x,\mathcal{D}_t)(F\cdot G)\,=\,\sum_{k=1}^N\mathcal{P}(\mathcal{D}_x,\mathcal{D}_t)(f_{k}\cdot G)e_k,\label{vectorDerivative}
\end{equation}
where $\mathcal{P}=\mathcal{P}(X,T)$ is a polynomial. The components $\mathcal{P}(\mathcal{D}_x,\mathcal{D}_t)(f_{k}\cdot G)$ are calculated using definition (\ref{HirotaD}). The mathematic foundation and properties of the vector Hirota derivative (\ref{vectorDerivative}) are discussed in appendix A. Here are some relevant examples using substitution (\ref{subst}):
\begin{eqnarray}
    w_t&=&\frac{\mathcal{D}_t(F\cdot G)}{G^2},\quad w_x=\frac{\mathcal{D}_x(F\cdot G)}{G^2},\\
    w_{xxx}&=&\frac{\mathcal{D}_x^3(F\cdot G)}{G^2}-3\frac{\mathcal{D}_x^2(G\cdot G)\mathcal{D}_x(F\cdot G)}{G^4}.
\end{eqnarray}
Introducing these examples in the cmKdV equations (\ref{V2cmKdV}), we get a set of two vector Hirota bilinear equations:
\begin{eqnarray}
    (\mathcal{D}_t+\mathcal{D}_x^3)(F\cdot G)&=&0,\label{HEq1}\\
    \mathcal{D}_x^2(G\cdot G)-2 F^t\mathcal{E}F&=&0\label{HEq2}.
\end{eqnarray}

\medskip

It is important to emphasize that, in contrast with standard component-wise approaches, the present formulation preserves the vector structure throughout the bilinearization procedure and is basis independent (see appendix A).  In particular, the nonlinear coupling term $F^t \mathcal{E} F$ appears naturally as a quadratic form, highlighting the role of the coupling matrix in the interaction dynamics. This provides a more compact and structurally consistent framework for the analysis of multi-component soliton solutions.

 \section{Vector Multi-Soliton Solutions}  To find soliton solutions of the system of bilinear equations (\ref{HEq1}) and (\ref{HEq2}), one needs to find the constant solutions of these equations $F_0$ and $G_0$. Once these solutions are found, we expand the unknown functions $F$ and $G$ around $F_0$ and $G_0$. These expansions are called $\epsilon$-expansions and we note
 \begin{equation}
     F=F_0+\epsilon F_1+\epsilon^2F_2+\cdots\quad \mbox{and}\quad G=G_0+\epsilon G_1+\epsilon^2G_2+\cdots.
 \end{equation}
Here we will assume without any lost of generality that the ground state solution for $G$ is given by $G_0\equiv 1$. Furthermore, in soliton theory, these seemingly infinite expansions are actually finite assuming that each soliton be modeled by an independent exponential function. As a consequence, no convergence problem arises and the obtained solutions are exact and not approximations.

In this case, the ground state solutions $F_0$ and $G_0\equiv1$ satisfy the constraint:
\begin{equation}
     F_0^t\mathcal{E}F_0=0.
    \label{groundstate}
\end{equation}

In the focusing and defocusing cases, where $\mathcal{E}=\pm I_N$, the ground state constraint reduces to $\vert F_0\vert^2=F_0^tF_0=0$ which admits the unique trivial solution $F_0=0$. In this particular case, the soliton solutions are represented as depressions or lumps on a uniform zero background. In general, this is not the case.  Indeed, for $N\geq 2$ and $\mathcal{E}\neq\pm I_N$, we may assume, without loss of generality, that there exist two integers $n_1\geq1$ and $n_2\geq1$ such that $n_1+n_2=N$ and $\mathcal{E}=\mbox{diag}(I_{n_1},-I_{n_2})$. In this general case, the ground state constraint (\ref{groundstate}) still admits the trivial solution $F_0$, but also admits non-zero vector solution. The latter case will be exploited in a later section as an opening for future perspectives and will demonstrate the existence of  kink-like solutions of the cmKdV system.

In what follows, we will assume $F_0=0$ and the $\epsilon-$expansions for the unknown functions $F$ and $G$:
\begin{equation}
    F=\epsilon F_1+\epsilon^3 F_3+\epsilon^5 F_5+\cdots\quad\mbox{and}\quad G=1+\epsilon^2G_2+\epsilon^4 G_4+\cdots,
\end{equation}
where $F_{2k-1}=F_{2k-1}(x,t)$ are real $N-$components vector functions and $G_{2k}=G_{2k}(x,t)$ are real-valued scalar functions for all integer $k\geq 1$.

\subsection{The vector one-soliton solution}

\begin{figure}[h]
    \centering
    \includegraphics[width=1\textwidth]{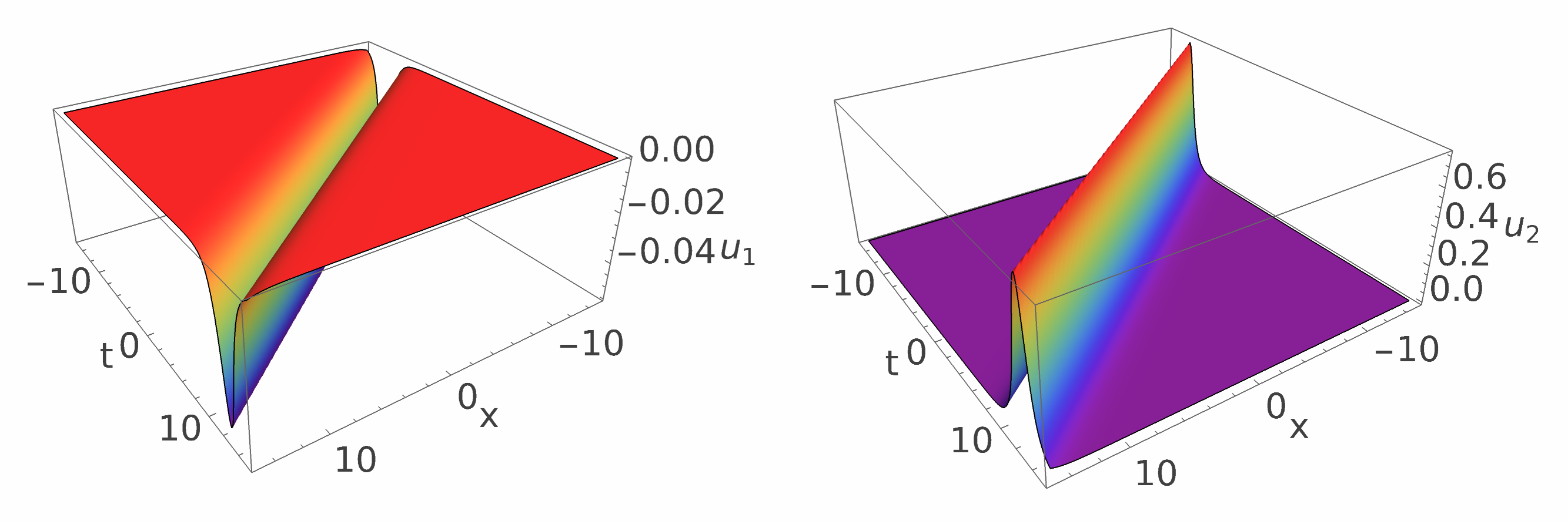}
    \caption{One-soliton vector solution illustrating the component structure of the wave profile for $u_1$ and $u_2$ given in equations (\ref{u1u21sol}).}
    \label{1SolitoncmKdv}
\end{figure}

For the traveling wave or, equivalently, the one-soliton solution, we assume the finite $\epsilon$-expansions for the functions $F$ and $G$:
\begin{equation}
    F=\epsilon F_1\quad \mbox{and}\quad G=1+\epsilon^2 G_2.\label{travelFG}
\end{equation}
 Introducing these functional expansions in equations (\ref{HEq1}) and (\ref{HEq2}), we obtain a system of four partial differential equations (PDEs) to solve by equating to zero each power of $\epsilon$:
\begin{eqnarray}
    F_{1,t}+F_{1,xxx}=0,\quad G_{2,xx}-F_1^t\mathcal{E}F_1=0,\\
    (\mathcal{D}_t+\mathcal{D}_x^3)(F_1\cdot G_2)=0,\quad \mathcal{D}_x^2(G_2\cdot G_2)=0.
\end{eqnarray}
For the traveling wave solution, we take
\begin{equation}
    F_1=\zeta \,e^{\eta}\quad \mbox{and}\quad G_2=\alpha\, e^{2\eta},\label{F1G2}
\end{equation}
where $\zeta$ is an arbitrary vector of $\mathbb{R}^N$ and $\eta=\kappa\, x+\omega\, t+\varphi$. Upon substitution, we get 
\begin{equation}
    \omega+\kappa^3=0\quad\mbox{and}\quad\alpha=\frac{\zeta^t\mathcal{E}\zeta}{4\kappa^2}.\label{condition1S}
\end{equation}

For example, let us consider the focusing case ($\mathcal{E}=I_N$), $N=2$ and assume that the matrix $A$ in equation (\ref{cmKdV}) is positive-definite. Let us say
\begin{equation}
    A=\begin{pmatrix}
        2&1\\
        1&2
    \end{pmatrix}
\end{equation}
The eigenvalues of $A$ are given by $\lambda_1=1$ and $\lambda_2=3$. The orthogonal matrix $P$ is given by
\begin{equation}
    P=\frac{1}{\sqrt{2}}\begin{pmatrix}
        1&1\\
        -1&1
    \end{pmatrix}.
\end{equation}
So, considering $u^t=(u_1\,\, u_2)$ and $w^t=(w_1\,\, w_2)$, we get
\begin{equation}
    u_1=\frac{1}{\sqrt2}\left(w_1-\frac{w_2}{\sqrt{3}}\right)\quad\mbox{and}\quad u_2=\frac{1}{\sqrt2}\left(w_1+\frac{w_2}{\sqrt{3}}\right).\label{u1u2}
\end{equation}

In Figure~\ref{1SolitoncmKdv}, we give the functions $u_1$ and $u_2$ for the one-soliton solution and the following choices of parameters: $\zeta^t=(1\,\,\,2)$, $\kappa=\frac{\sqrt{5}}{2}$, $\epsilon=1$ and $\varphi=0.$ In this particular case, we get explicitly
\begin{equation}
    u_1=\frac{1}{2\sqrt2}\left(1-\frac{2}{\sqrt3}\right)\mbox{sech}(\eta)\quad \mbox{and}\quad u_2=\frac{1}{2\sqrt2}\left(1+\frac{2}{\sqrt3}\right)\mbox{sech}(\eta),\label{u1u21sol}
\end{equation}
for $\eta=\frac{\sqrt5}{2}(x-\frac{5}{4}t).$

Figure~\ref{1SolitoncmKdv} illustrates the vector one-soliton solution through its component profiles $u_1$ and $u_2$. Although the solution retains the characteristic $\mathrm{sech}$ profile of the scalar mKdV soliton, the amplitudes of the individual components are modulated by the coupling structure encoded in the matrix $A$. 

In particular, the transformation from the diagonalized variables to the physical components induces an asymmetry between $u_1$ and $u_2$, reflecting the underlying coupling between modes. This highlights a key feature of the vector formulation: even in the simplest one-soliton regime, the wave profile is not a trivial duplication of scalar solutions but rather a coupled structure determined by the eigenvalues and eigenvectors of the interaction matrix.

This demonstrates that the vector approach naturally incorporates mode mixing effects at the level of exact solutions.

\subsection{The vector two-soliton solution}

For the two soliton solution, we assume the following $\epsilon$-expansions for the functions $F$ and $G$:
\begin{equation}
    F=\epsilon F_1+\epsilon^3 F_3\quad \mbox{and}\quad G=1+\epsilon^2G_2+\epsilon^4G_4.\label{2SForm}
\end{equation}
Using a similar procedure as for the traveling wave solution, we get a set of PDEs to solve for the component functions of $F$ and $G$. For a two-soliton solution, we assume
\begin{equation}
    F_1=\zeta_1 e^{\eta_1}+\zeta_2 e^{\eta_2},\label{F12S}
\end{equation}
where $\zeta_1$ and $\zeta_2$ are arbitrary constant vectors of $\mathbb{R}^N$, and $\eta_j=\kappa_j\, x+\omega_j\, t+\varphi_j$ for $j=1,2$. The PDE for $F_1$ being linear, we get, from the superposition principle, the dispersion relations:
\begin{equation}
    \omega_j+\kappa_j^3=0.\label{disperse}
\end{equation}
The function $G_2$ is obtained from the equation associated to $\epsilon^2$. We get
\begin{equation}
    G_2=\alpha_{11}e^{2\eta_1}+2\alpha_{12}e^{\eta_1+\eta_2}+\alpha_{22}e^{2\eta_2}\label{G22S}
\end{equation}
where
\begin{equation}
    \alpha_{mn}=\alpha_{nm}=\frac{\zeta_m^t\mathcal{E}\zeta_n}{(\kappa_m+\kappa_n)^2}.\label{alphamn}
\end{equation}
For $F_3$, we make the ansatz
\begin{equation}
    F_3=\chi_{1}^{22}e^{\eta_1+2\eta_2}+\chi_2^{11} e^{\eta_2+2\eta_1}.\label{F32S}
\end{equation}
We find
\begin{equation}
    \chi_m^{nn}=\zeta_m\alpha_{nn}\beta_{12}^2,\quad \beta_{12}=\frac{\kappa_1-\kappa_2}{\kappa_1+\kappa_2}.
\end{equation}
For $G_4$, the $\epsilon^4$ equation yields
\begin{equation}
    G_4=A_{12}\, e^{2\eta_1+2\eta_2},\quad A_{12}=\alpha_{11}\alpha_{22}\beta_{12}^4.\label{G42S}
\end{equation}
The above components are sufficient to solve the set of bilinear equations (\ref{HEq1}) and (\ref{HEq2}), thus obtaining an exact two-soliton solution. In appendix B, we show that our proposed two-soliton solution is obtained for the integrability condition (\ref{2SolitonIntegra}). This integrability condition holds under the dispersion relations (\ref{disperse}) which is coherent with partial Hirota integrability. 

We consider the same example as for the one-soliton solution where the component functions $u_1$ and $u_2$ are given by equation (\ref{u1u2}). 

\begin{figure}[h]
    \centering
    \includegraphics[width=\textwidth]{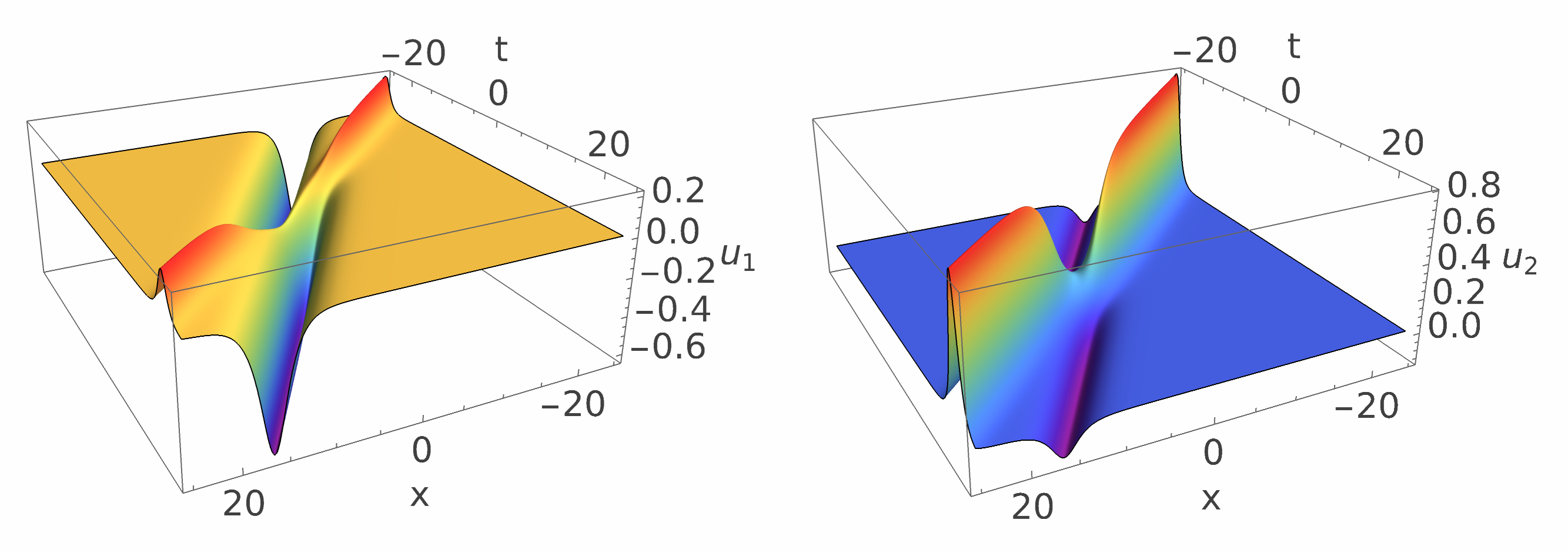}
    \caption{Two-soliton interaction showing a mixed bright-dark structure and elastic collision behavior. The left-hand side is for the component $u_1$ and the right-hand figure is for $u_2$ given in equations (\ref{u1u2}) for the parameter choices: $\epsilon=1$, $\varphi_1=\varphi_2=0$, $\kappa_1=0.8$, $\kappa_2=1$, $\zeta_1^t=(-1\,\,\, 1)$ and $\zeta_2^t=(1\,\,\, 1)$.}
    \label{2SolitoncmKdv}
\end{figure}

In Figure~\ref{2SolitoncmKdv}, we show the components $u_1$ and $u_2$ for a vector two-soliton solution and for the following choice of parameters : $\epsilon=1$, $\varphi_1=\varphi_2=0$, $\kappa_1=0.8$, $\kappa_2=1$, $\zeta_1^t=(-1\,\,\, 1)$ and $\zeta_2^t=(1\,\,\, 1)$. 

The collisions exhibit a mixed bright-dark structure, where different components display distinct localized behaviors during the interaction process.

This feature is a direct consequence of the vector coupling, which allows energy redistribution between components during the collision. Despite this complex internal structure, the interaction remains elastic: the solitons recover their original shapes and velocities after collision, which is a hallmark of integrable soliton dynamics.

Compared to the scalar case, where soliton interactions are purely scalar and symmetric, the vector formulation reveals additional interaction channels governed by the coupling matrix. This leads to richer dynamical patterns while preserving integrability.

\subsection{The vector three-soliton solution}

As mentioned in the introduction, the existence of a three-soliton solution was considered to be a condition for complete integrability of systems. It was already shown that the system of coupled mKdV equations (\ref{cmKdV}) is completely integrable using the inverse scattering transform method. It should then be expected that a three soliton solution should exist for this system.

We assume these $\epsilon-$expansions for the unknown functions $F$ and $G$:
\begin{equation}
    F=\epsilon F_1+\epsilon^3F_3+\epsilon^5F_5\quad \mbox{and}\quad G=1+\epsilon^2G_2+\epsilon^4G_4+\epsilon^6G_6.
\end{equation}
For the three-soliton solution, we take
\begin{equation}
    F_1=\zeta_1e^{\eta_1}+\zeta_2e^{\eta_2}+\zeta_3e^{\eta_3},
\end{equation}
where $\zeta_j$ are arbitrary constant vectors of $\mathbb{R}^N$ and $\eta_j=\kappa_jx+\omega_jt+\varphi_j$ for $j=1,2,3$. The constant $\kappa_j$ and $\omega_j$ satisfy the usual dispersion relation (\ref{disperse}). The other components are given explicitly by
\begin{eqnarray}
    G_2&=&\alpha_{11}e^{2\eta_1}+\alpha_{22}e^{2\eta_2}+\alpha_{33}e^{2\eta_3}+2\alpha_{12}e^{\eta_1+\eta_2}+2\alpha_{13}e^{\eta_1+\eta_3}+2\alpha_{23}e^{\eta_2+\eta_3},\\
    F_3&=&\chi_1^{22}e^{\eta_1+2\eta_2}+\chi_1^{33}e^{\eta_1+2\eta_3}+\chi_2^{11}e^{\eta_2+2\eta_1}+\chi_2^{33}e^{\eta_2+2\eta_3}\nonumber\\&+&\chi_3^{11}e^{\eta_3+2\eta_1}
    +\chi_3^{22}e^{\eta_3+2\eta_2}+2(\chi_1^{23}+\chi_2^{13}+\chi_3^{12})e^{\eta_1+\eta_2+\eta_3},\\
    G_4&=&A_{12}e^{2\eta_1+2\eta_2}+A_{13}e^{2\eta_1+2\eta_3}+A_{23}e^{2\eta_2+2\eta_3}+2C_1^{23}e^{2\eta_1+\eta_2+\eta_3}\nonumber\\
    &+&2C_2^{13}e^{\eta_1+2\eta_2+\eta_3}+2C_3^{12}e^{\eta_1+\eta_2+2\eta_3},\label{G4}\\
    F_5&=&\zeta_1^{23}e^{\eta_1+2\eta_2+2\eta_3}+\zeta_2^{13}e^{2\eta_1+\eta_2+2\eta_3}+\zeta_3^{12}e^{2\eta_1+2\eta_2+\eta_3},\\
    G_6&=&C_{123}e^{2\eta_1+2\eta_2+2\eta_3},
\end{eqnarray}
where the constants $\alpha_{mn}$ are defined in equation (\ref{alphamn}) and are such that $\alpha_{mn}=\alpha_{nm}.$ For the other constants, we get explicitly
\begin{eqnarray}
    \chi_j^{mn}&=&\zeta_j\alpha_{mn}\beta_{mj}\beta_{nj},\quad \beta_{mj}=\frac{\kappa_m-\kappa_j}{\kappa_m+\kappa_j}\label{betaterms},\\
    C_j^{mn}&=&\alpha_{jj}\alpha_{mn}\beta_{mj}^2\beta_{nj}^2,\quad A_{jm}=C_j^{mm},\\
    \zeta_j^{mn}&=&\zeta_j\alpha_{mm}\alpha_{nn}\beta_{mj}^2\beta_{nj}^2\beta_{mn}^4,\\
    C_{123}&=&\alpha_{11}\alpha_{22}\alpha_{33}\beta_{12}^4\beta_{13}^4\beta_{23}^4.
\end{eqnarray}

We have constructed all the components of the unknown functions $F$ and $G$. We have shown the existence of a three-soliton solution for the coupled system of mKdV equations (\ref{cmKdV}). This result is consistent with the integrability of the cmKdV system established through the inverse scattering transform and confirms that the vector bilinear formulation correctly reproduces the three-soliton structure.

To formally link our paper to the outstanding work of Hietarinta \cite{Hietarinta1,Hietarinta2,Hietarinta3,Hietarinta4}, we need to establish the three-soliton Hirota integrability condition. We show, in appendix B, that the three-soliton Hirota integrability condition is given by equation (\ref{3SSIn}) which holds whenever the dispersion relations (\ref{disperse}) are satisfied. Showing formally that the cmKdV (\ref{cmKdV}) is integrable in the Hirota sense.

\begin{figure}[h]
    \centering
    \includegraphics[width=1\textwidth]{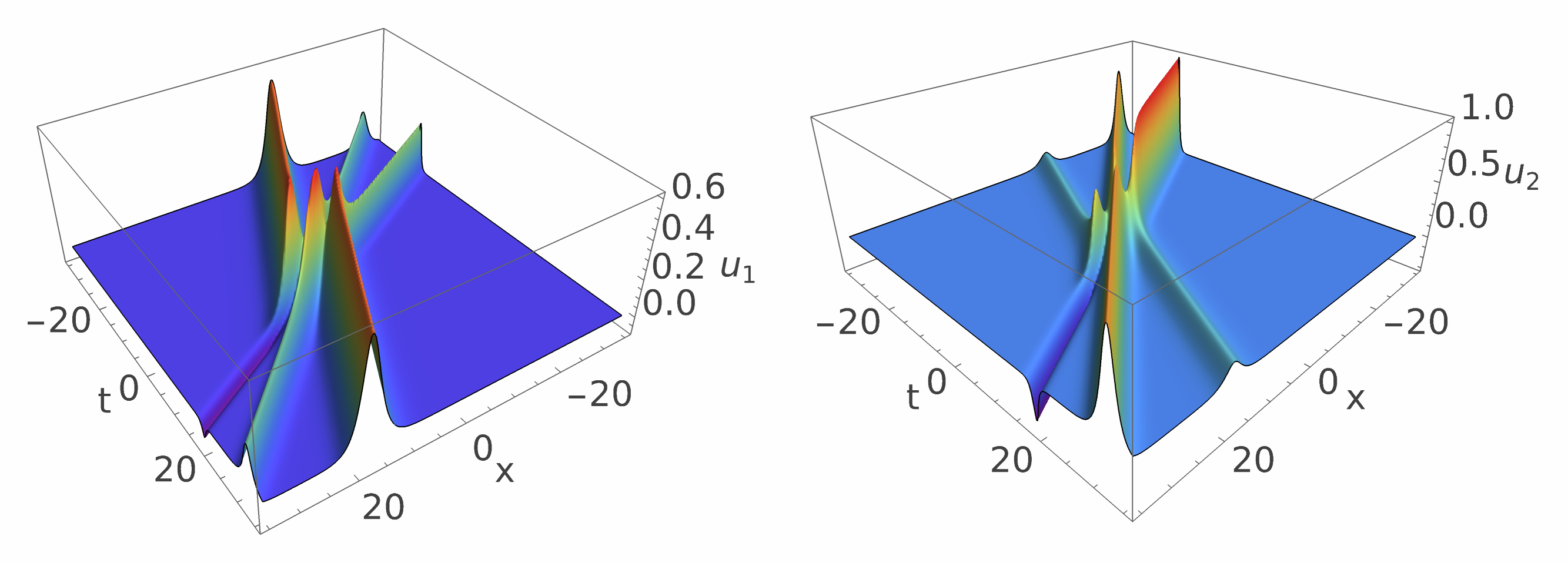}
    \caption{Three-soliton interaction illustrating complex vector coupling dynamics and multi-channel interactions.  The left-hand side is for the component $u_1$ and the right-hand figure is for $u_2$ given in equations (\ref{u1u2}) for the parameter choices: $\epsilon=1$, $\kappa_1=1$, $\kappa_2=0.7$, $\kappa_3=1.3$, $\varphi_1=\varphi_2=\varphi_3=0$, $\zeta_1^t=\zeta_3^t=(1\,\,\,1)$ and $\zeta_2^t=(1\,\,\,-1)$.}
    \label{3SolitoncmKdv}
\end{figure}

We consider the same example as for the vector one and two-soliton solutions where the components of the vector solution $u_1$ and $u_2$ are given in equation (\ref{u1u2}). 
In Figure~\ref{3SolitoncmKdv}, we give a representation of the functions $u_1$ and $u_2$ for a three-soliton solution and for the following choice of parameters: $\epsilon=1$, $\kappa_1=1$, $\kappa_2=0.7$, $\kappa_3=1.3$, $\varphi_1=\varphi_2=\varphi_3=0$, $\zeta_1^t=\zeta_3^t=(1\,\,\,1)$ and $\zeta_2^t=(1\,\,\,-1)$.

The three-soliton configuration illustrates the complex interaction patterns arising in the vector case. The interaction pattern displays multiple overlapping collision events, corresponding to different pairwise and collective interaction channels between the solitons.

The persistence of well-defined soliton structures after these interactions confirms the elastic nature of collisions and is consistent with the existence of a three-soliton solution. In the context of Hirota's method, this provides a direct manifestation of the integrability of the system at the vector level.

Moreover, the vector formulation reveals that these interactions are not merely superpositions of scalar processes, but involve coupled dynamics across components, leading to a richer structure of multi-soliton interactions.

\section{The one-soliton solution for non-trivial background} In the previous section, the soliton solutions were constructed from $\epsilon-$expansions around the ground state solutions $(F_0,G_0)=(0,1)$. Indeed, we have taken the trivial solution of the ground state constraint (\ref{groundstate}). However, as discussed previously, for $N\geq 2$ and $\mathcal{E}\neq \pm I_N$, the ground state equation (\ref{groundstate}) admits non-trivial solution $F_0$. As a simple example, if $\mathcal{E}=\mbox{diag}(1,-1)$, then $F_0^t=(1\,\,\, 1)$ is a solution of the constraint yielding soliton solution on a non-zero background. For $\mathcal{E}\neq \pm I_N$ and for future considerations, we may thus find at least two non-trivial vectors $v_1\neq v_2$ solutions to the equation
\begin{equation}
    v^t\mathcal{E}v=0.\label{MatrixCo}
\end{equation}
Taking the previous example, $v_1^t=(1\,\,\,1)$ and $v_2^t=(1\,\,\, -1)$ are two distinct solutions of the ground state constraint for $\mathcal{E}=\mbox{diag}(1,-1)$.

To construct a vector one-soliton solution, we assume that the unknown functions $F$ and $G$ in the bilinear equations (\ref{HEq1}) and (\ref{HEq2}) have the forms:
\begin{equation}
    F=F_0+\epsilon F_1\quad \mbox{and}\quad G=1+\epsilon G_1.
\end{equation}
We find that
\begin{equation}
    F_1=v\, e^{\eta}\quad \mbox{and}\quad G_1=\beta\, e^{\eta},
\end{equation}
where $\eta=\kappa x+\omega t+\varphi$, $v\neq F_0$ is a non-trivial vector solution of equation (\ref{MatrixCo}) and $\beta$ is a real constant. We get the usual dispersion relation $\omega+\kappa^3=0$ and 
\begin{equation}
    \beta=\frac{2F_0^t\mathcal{E}v}{\kappa^2}.
\end{equation}

\medskip

\begin{figure}[h]
    \centering
    \includegraphics[width=1\textwidth]{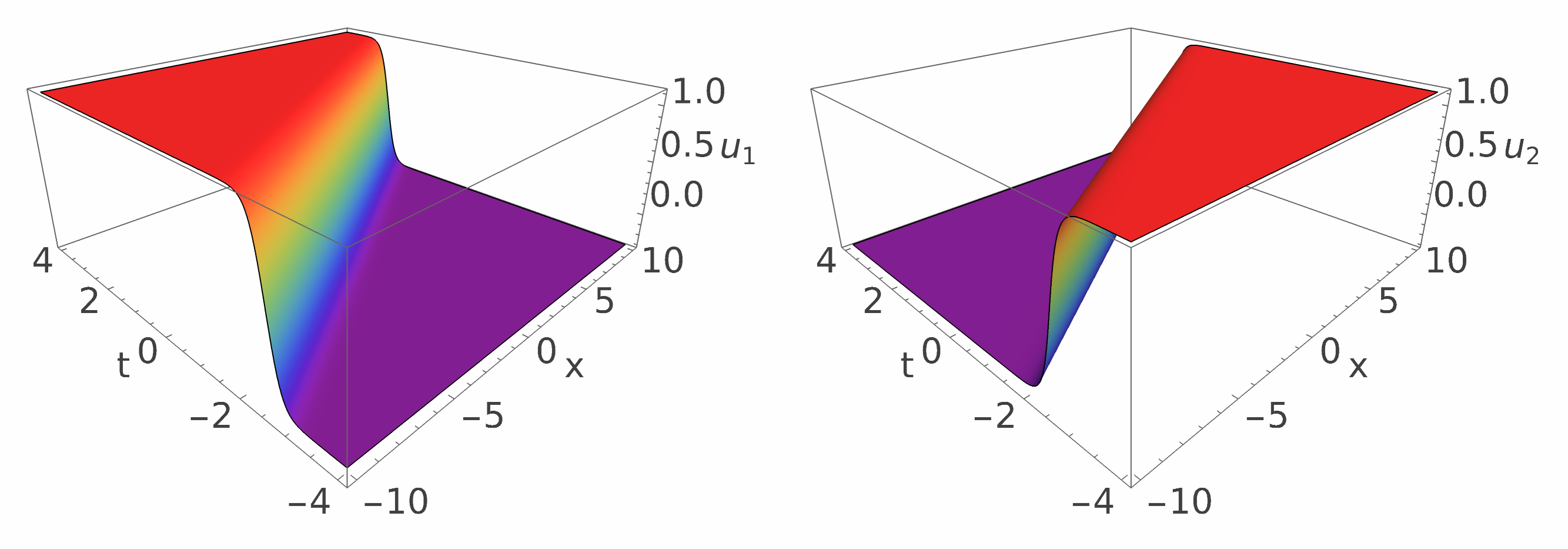}
    \caption{One-soliton vector solution on a non-zero background for the component structures $u_1$ and $u_2$ given in Equation (\ref{NonZero1}).}
    \label{1SolitoncmKdvNZero}
\end{figure}

This construction provides a first step toward the analysis of vector solitons on nontrivial backgrounds, which are known to exhibit richer dynamics than standard zero-background solitons.

Let us consider an example. Suppose that $N=2$ and that the matrix $A$ of the original system (\ref{cmKdV}) is given by
\begin{equation}
    A=\begin{pmatrix}
        1&2\\
        2&1
    \end{pmatrix}.
\end{equation}
In this case, we get the eigenvalues $\lambda_1=3$ and $\lambda_2=-1$. The matrix $P$ constructed from the eigenvectors is 
\begin{equation}
    P=\frac{1}{\sqrt{2}}\begin{pmatrix}
        1&1\\
        1&-1
    \end{pmatrix}.
\end{equation}
The vector solution $u^t=(u_1\,\,\, u_2)$ is obtained from the vector $w^t=(w_1\,\,\,w_2)$. We get
\begin{equation}
    u_1=\frac{1}{\sqrt2}\left(\frac{w_1}{\sqrt3}+w_2\right)\quad \mbox{and}\quad u_2=\frac{1}{\sqrt2}\left(\frac{w_1}{\sqrt3}-w_2\right).
\end{equation}
One can take $F_0^t=(1\,\,\,1)$, $v^t=(1\,\,\,-1)$, $\epsilon=1$, $\kappa=2$ and $\varphi=0$. In this special case, we get
\begin{equation}
    u_1=\frac{1}{\sqrt2}\left(\frac{1}{\sqrt3}-\tanh\left(\frac{\eta}{2}\right)\right)\quad \mbox{and}\quad u_2=\frac{1}{\sqrt2}\left(\frac{1}{\sqrt3}+\tanh\left(\frac{\eta}{2}\right)\right).\label{NonZero1}
\end{equation}

In Figure~\ref{1SolitoncmKdvNZero}, we give the functions $u_1$ and $u_2$ for the one-soliton solution on a non-zero background described by Equation (\ref{NonZero1}).

Figure~\ref{1SolitoncmKdvNZero} presents a one-soliton solution on a non-zero background arising from a nontrivial vector ground state. In contrast to the standard $\mathrm{sech}$-type solitons obtained in the zero-background case, the solution exhibits a $\tanh$-type profile, characteristic of dark or kink-like structures.

This behavior originates from the existence of nontrivial solutions to the constraint $F_0^t \mathcal{E}  F_0 = 0$ in the case of indefinite coupling. Such solutions have no analogue in the scalar mKdV equation and are a direct consequence of the vector nature of the system. Indeed, in the scalar case, we distinguish only two cases: focusing and defocusing.

The result highlights a fundamental advantage of the vector bilinear formulation: it naturally captures nonlinear excitations on nontrivial backgrounds, opening the way to the study of more complex structures such as dark solitons, domain walls, and mixed bright–dark configurations in multi-component systems.

\vspace{0.5cm}

These examples illustrate that the vector bilinear formulation not only reproduces known integrable features but also reveals new structural properties of multi-component soliton dynamics, including component-dependent profiles, energy redistribution during interactions, and the emergence of nontrivial background states.

\section{Conclusion} In this work, we have developed a vector reformulation of Hirota's bilinear formalism for the coupled modified Korteweg--de Vries (cmKdV) system with a real symmetric coupling matrix. By expressing both the bilinear equations and their solutions directly at the vector level, the proposed approach preserves the intrinsic structure of the coupled model and provides a compact, structurally consistent alternative to standard component-wise formulations. Within this framework, we have constructed explicit one-, two-, and three-soliton solutions in closed vector form. The recovery of the three-soliton solution confirms that the vector bilinear formulation fully captures the integrable structure of the cmKdV system, in agreement with results obtained from the inverse scattering transform. More generally, the approach provides a unified description of focusing, defocusing, and mixed-sign regimes, where the role of the coupling matrix appears explicitly through quadratic forms governing the nonlinear interactions.

A key feature of the present formulation is the natural emergence of nontrivial vector ground states in the case of indefinite coupling. This property leads to the existence of soliton solutions on non-zero backgrounds, which have no direct analogue in the scalar mKdV equation. This result highlights the structural richness of multi-component systems and demonstrates that the vector bilinear approach is not only a reformulation, but also a framework capable of revealing new classes of nonlinear excitations.

Beyond the explicit constructions presented here, the vector bilinear formalism offers a promising foundation for the systematic study of multi-component integrable systems. In particular, it would be of interest to extend this approach to multi-soliton solutions on nontrivial backgrounds, as well as to other classes of exact solutions such as rogue waves, rational solutions, and periodic structures. More broadly, the methodology developed in this work may be applied to other coupled integrable models, providing a unified and structure-preserving framework for the analysis of vector nonlinear wave dynamics.

\subsection*{Acknowledgements}

We thank the reviewer for their insightful comments and constructive suggestions, which have significantly improved the quality of our manuscript.

\appendix

\section{Mathematic foundation of the vector Hirota derivative}In this appendix, we discuss the mathematical foundation and properties of vector Hirota derivative definition (\ref{vectorDerivative}).
\begin{enumerate}
    \item[(1)] The vector Hirota derivative (\ref{vectorDerivative}) is a bilinear form. For all $\lambda\in \mathbb{R}$, real- valued vector functions $F_1$ and $F_2$, and real-valued scalar functions $G_1$ and $G_2$, we have
    \begin{enumerate}
        \item[$\bullet$] $\mathcal{P}((F_1+\lambda F_2)\cdot G)=\mathcal{P}(F_1\cdot G)+\lambda \,\mathcal{P}( F_2\cdot G)$
        \item[$\bullet$] $\mathcal{P}(F\cdot (G_1+\lambda G_2))=\mathcal{P}(F\cdot G_1)+\lambda \,\mathcal{P}( F\cdot G_2)$
    \end{enumerate}
    \textit{Proof:} Let us write the vector functions $F_1$ and $F_2$ in component form:
    \begin{equation}
        F_1(x,t)=\sum_{k=1}^Nf_{1,k}(x,t)e_k \quad \mbox{and}\quad F_2(x,t)=\sum_{k=1}^Nf_{2,k}(x,t)e_k.
    \end{equation}
    We have
    \begin{eqnarray*}
        \mathcal{P}((F_1+\lambda F_2)\cdot G)&=&\sum_{k=1}^N\mathcal{P}((f_{1,k}+\lambda f_{2,k})\cdot G)e_k\\
        &=&\sum_{k=1}^N\left(\mathcal{P}(f_{1,k}\cdot G)+\lambda \mathcal{P}( f_{2,k}\cdot G)\right)e_k\\
        &=&\sum_{k=1}^N\mathcal{P}(f_{1,k}\cdot G)e_k+\lambda\sum_{k=1}^N\mathcal{P}(f_{2,k}\cdot G)e_k\\
        &=&\mathcal{P}(F_1\cdot G)+\lambda \mathcal{P}( F_2\cdot G).
    \end{eqnarray*}
    The second equality is obtained using the bilinear property of the classical Hirota derivative (\ref{HirotaD}). This shows the first property. For the second one, we write, as for $F_1$ and $F_2$, the vector function $F$ in components. We have
    \begin{eqnarray}
        \mathcal{P}(F\cdot (G_1+\lambda G_2))&=&\sum_{k=1}^N\mathcal{P}(f_k\cdot(G_1+\lambda G_2))e_k\\
        &=&\sum_{k=1}^N\left({P}(f_k\cdot G_1)+\lambda {P}(f_k\cdot G_2)\right)e_k\\
        &=&\sum_{k=1}^N{P}(f_k\cdot G_1)e_k+\lambda\sum_{k=1}^N{P}(f_k\cdot G_2)e_k\\
        &=&\mathcal{P}(F\cdot G_1)+\lambda \mathcal{P}(F\cdot  G_2)
    \end{eqnarray}
    The second equality is obtained from the bilinear property of the classical Hirota derivative (\ref{HirotaD}). 
    \item[(2)] The vector Hirota derivative (\ref{vectorDerivative}) is basis-independent.\\
    Let us consider $\mathcal{B}_1=\{e_1,e_2,\cdots,e_N\}$ and $\mathcal{B}_2=\{\tilde{e}_1,\tilde{e}_2,\cdots,\tilde{e}_N\}$ two basis of $\mathbb{R}^N$. For all $k$, we have
    \begin{equation}
        \tilde{e}_k=\sum_{j=1}^Na_{jk}e_j,
    \end{equation}
    for unique real constants $a_{jk}$. We may thus write the $N$-component vector function $F$ using these two basis. We have
    \begin{eqnarray*}
        F(x,t)=\sum_{j=1}^Nf_j(x,t)e_i&=&\sum_{k=1}^N\tilde{f_k}(x,t)\tilde{e}_k\\
        &=&\sum_{k=1}^N\tilde{f}_k(x,t)\sum_{j=1}^Na_{jk}e_j\\
        &=&\sum_{j=1}^N\left(\sum_{k=1}^Na_{jk}\tilde{f}_k(x,t)\right)e_j
    \end{eqnarray*}
    By uniqueness of basis representation, we deduce that
    \begin{equation}
        f_j(x,t)=\sum_{k=1}^Na_{jk}\tilde{f}_k(x,t),
    \end{equation}
    for all integer $1\leq j\leq N$. We deduce, using the bilinear properties of the  Hirota derivative (\ref{HirotaD}) and the definition of the vector Hirota derivative (\ref{vectorDerivative}), that
    \begin{eqnarray*}
        \mathcal{P}(F\cdot G)&=&\sum_{j=1}^N\mathcal{P}(f_j\cdot G)e_j\\
        &=&\sum_{j=1}^N\mathcal{P}\left(\left(\sum_{k=1}^Na_{jk}\tilde{f}_k\right)\cdot G\right)e_j\\
        &=&\sum_{j=1}^N\sum_{k=1}^Na_{jk}\mathcal{P}(\tilde{f}_k\cdot G)e_j\\
       &=& \sum_{k=1}^N\mathcal{P}(\tilde{f}_k\cdot G)\sum_{j=1}^Na_{jk}e_j\\
       &=&\sum_{k=1}^N\mathcal{P}(\tilde{f}_k\cdot G)\tilde{e}_k
    \end{eqnarray*}
    Showing that the definition of the vector Hirota derivative (\ref{vectorDerivative}) is independent of the chosen $\mathbb{R}^N$ basis.
\end{enumerate}

\section{The soliton integrability condition}In this appendix, we establish the three-soliton Hirota integrability conditions. To obtain these conditions, let us write the bilinear equations (\ref{HEq1}) and (\ref{HEq2}) as
\begin{eqnarray}
    \mathcal{P}_1(\mathcal{D}_x,\mathcal{D}_t)(F\cdot G)=0 \quad \mbox{and}\quad \mathcal{P}_2(\mathcal{D}_x,\mathcal{D}_t)(G\cdot G)=2F^{t}\mathcal{E}F,
\end{eqnarray}
where $\mathcal{P}_1(X,T)=T+X^3$ and $\mathcal{P}_2(X,T)=X^2$. For convenience, we consider the vector $\vec{\eta}_j=(\kappa_j,\omega_j)$ to refer to the exponential arguments $\eta_j=\kappa_j x+\omega_j t+\varphi_j=\vec{\eta}_j\cdot \vec{x}+\varphi_j$ for $\vec{x}=(x,t)$.
\begin{enumerate}
    \item[(1)] The one-soliton solution condition.\\
    The system of bilinear equations (\ref{HEq1}) and (\ref{HEq2}) admits a vector one-soliton solution for the functions $F$ and $G$ constructed in Section IV.A if the dispersion relation:
    \begin{equation}
        \mathcal{P}_1(\vec{\eta})=0\quad \Longleftrightarrow\quad \omega+\kappa^3=0
    \end{equation}
    is satisfied.
    \item[(2)] The two-soliton conditions. \\
    The system of bilinear equations (\ref{HEq1}) and (\ref{HEq2}) admits a vector two-soliton solution for the functions $F$ and $G$ given in Section IV.B if we have the dispersion relations $\mathcal{P}
    _1(\vec{\eta}_1) = \mathcal{P}_1(\vec{\eta}_2) = 0$ and
    \begin{equation}
        \mathcal{P}_2(\vec{\eta}_p+\vec{\eta}_q)\mathcal{P}_1(2\vec{\eta}_p-\vec{\eta}_q)-\mathcal{P}_2(\vec{\eta}_p-\vec{\eta}_q)\mathcal{P}_1(2\vec{\eta}_p+\vec{\eta}_q)=0,\label{2SolitonIntegra}
    \end{equation}
    which holds on the two-soliton dispersion manifold $\mathcal{M}_2=\{(\vec{\eta_1},\vec{\eta_2})\, :\,\mathcal{P}_1(\vec{\eta}_1)=\mathcal{P}_1(\vec{\eta}_2)=0\}.$ This last condition shows that the system is partially integrable in the Hirota sense. Under these conditions, we have
    \begin{equation}
        \frac{\mathcal{P}_1(2\vec{\eta}_p-\vec{\eta}_q)}{\mathcal{P}_1(2\vec{\eta}_p+\vec{\eta}_q)}=\frac{\mathcal{P}_2(\vec{\eta}_p-\vec{\eta}_q)}{\mathcal{P}_2(\vec{\eta}_p+\vec{\eta}_q)}=\left(\frac{\kappa_p-\kappa_q}{\kappa_p+\kappa_q}\right)^2=\beta_{pq}^2..
    \end{equation}
    \item[(3)] The three-soliton Hirota integrability conditions.\\
    In Section IV.C, we have constructed a three-soliton solution showing implicitly the three-soliton integrability condition. To link this paper to the work of Hietarinta, we need to establish the Hirota integrability conditions which should hold on the three-soliton dispersion manifold $\mathcal{M}_3=\{(\vec{\eta_1},\vec{\eta_2},\vec{\eta_3})\,:\,\mathcal{P}(\vec{\eta}_j)=0,\, j=1,2,3\}$. \\
    On the two-soliton solution level, we find a first condition:
    \begin{equation}
    \frac{\mathcal{P}_2(\vec{\eta}_m+\vec{\eta}_q)\mathcal{P}_1(\vec{\eta}_p+\vec{\eta}_q-\vec{\eta}_m)+\mathcal{P}_2(\vec{\eta}_p+\vec{\eta}_q)\mathcal{P}_1(\vec{\eta}_m+\vec{\eta}_q-\vec{\eta}_p)}{\mathcal{P}_1(\vec{\eta}_m+\vec{\eta}_p+\vec{\eta}_q)}=\mathcal{P}_2(\vec{\eta}_m-\vec{\eta}_p),\label{Gen2SSI}
    \end{equation}
    which holds on $\mathcal{M}_3$. This condition generalizes the two-soliton condition (\ref{2SolitonIntegra}). Indeed, we retrieve the condition (\ref{2SolitonIntegra}) by taking $\vec{\eta_p}=\vec{\eta}_q$. We deduce, from condition (\ref{Gen2SSI}), the following identities:
    \begin{equation}
        \frac{\mathcal{P}_1(\vec{\eta}_m+\vec{\eta}_p-\vec{\eta}_q)}{\mathcal{P}_1(\vec{\eta}_m+\vec{\eta}_p+\vec{\eta}_q)}=\beta_{qm}\beta_{qp}.
    \end{equation}
    From the $\epsilon^5$ and $\epsilon^6$ equations (the three-soliton solution levels), we deduce the three-soliton Hirota integrability condition:
\begin{eqnarray}
    \beta_{mq}^2\frac{\mathcal{P}_1(2\vec{\eta}_p-2\vec{\eta}_q-\vec{\eta}_m)}{\mathcal{P}_1(\vec{\eta}_m+2\vec{\eta}_p+2\vec{\eta}_q)}+\beta_{mp}^2\frac{\mathcal{P}_1(2\vec{\eta}_q-2\vec{\eta}_p-\vec{\eta}_m)}{\mathcal{P}_1(\vec{\eta}_m+2\vec{\eta}_p+2\vec{\eta}_q)}+\beta_{pq}^4\frac{\mathcal{P}_1(2\vec{\eta}_p+2\vec{\eta}_q-\vec{\eta}_m)}{\mathcal{P}_1(\vec{\eta}_m+2\vec{\eta}_p+2\vec{\eta}_q)}\nonumber\\=\beta_{mp}^2\beta_{mq}^2\beta_{pq}^4,\label{3SSIn}
\end{eqnarray}
which holds on the three-soliton dispersion manifold $\mathcal{M}_3$. The three-soliton Hirota integrability condition (\ref{3SSIn}) generalizes the two-soliton partial integrability condition (\ref{2SolitonIntegra}). Indeed, projecting the manifold $\mathcal{M}_3$ onto the plane $\vec{\eta}_p=\vec{\eta}_m$ allows one to retrieve the manifold $\mathcal{M}_2$ and we can directly verify that the three-soliton integrability condition reduces to (\ref{2SolitonIntegra}).
\end{enumerate}

\label{lastpage}
\end{document}